\documentclass[11pt]{article}
\usepackage{amsmath}
\usepackage{amscd,amssymb}
\usepackage{hyperref}

\def\g{\gamma}
\newcommand{\be}{\begin{equation}}
\newcommand{\ee}{\end{equation}}
\newcommand{\bea}{\begin{eqnarray}}
\newcommand{\eea}{\end{eqnarray}}
\let\bm=\bibitem

\def\l{\lambda}

\def\t{\tau}
\def\th{\theta}

\def\4{{\sst{(4)}}}
\def\fr{\frac}

\begin{document}
\pagenumbering{roman}

\begin{titlepage}

\vspace{3.0cm}

\

\

\

\centerline{\Large \bf Deformations of Cosmological Solutions of D=11 Supergravity}
\vspace{1.5cm}

\centerline{Nihat Sadik Deger$^{1}$ and Ali Kaya$^{2}$}

\

\noindent $^1$ Dept. of Mathematics, Bogazici University, Bebek, 34342, Istanbul-Turkey \\
$^2$ Dept. of Physics, Bogazici University, Bebek, 34342, Istanbul-Turkey 

\

\noindent {\bf E-mails:} sadik.deger@boun.edu.tr , ali.kaya@boun.edu.tr

\vspace{1.5cm}

\centerline{\bf ABSTRACT}
\vspace{0.5cm}

We study Lunin-Maldacena deformations of cosmological backgrounds of D=11 supergravity which gives an easy way to generate solutions 
with nonzero 4-form flux starting from solutions of pure Einstein equations which possess at least three $U(1)$ isometries.  
We illustrate this on the vacuum S-brane solution from which the usual SM2-brane solution is obtained.
Applying the method again, one either gets the recently found S-brane system where contribution of the 
Chern-Simons term to field equations is non-zero or the SM2$\perp$SM2(0) intersection, depending on which $U(1)$ directions are used during 
the process. Repeated usage of the procedure gives rise to configurations with several overlapping S-branes some of which are new. 
We also employ this method to construct two more new solutions and  make comments about accelerating cosmologies that follow 
from such deformed solutions after compactification to (1+3)-dimensions.

\vspace{2cm}

\end{titlepage}

\pagenumbering{arabic}

\tableofcontents

\section{Introduction}

Using symmetries to find new solutions is quite an old and powerful idea. A string theory background 
which is independent of $d$-toroidal coordinates has an $O(d,d)$ T-duality symmetry (for a review see \cite{T-duality}). Exploiting this makes it  
possible to obtain new solutions starting from an existing one. This approach, also played an important role in the AdS/CFT context in 
understanding the gravity duals of deformed conformal field theories \cite{mal2}. If the ten dimensional gravity 
dual has a $U(1) \times U(1)$ symmetry in its geometry then such a solution can be deformed using the so called TsT transformation where a shift 
is sandwiched in between two T-dualities. For 11-dimensional solutions one just needs an extra $U(1)$ for the dimensional reduction to ten 
dimensions \cite{mal2}. After that, the TsT transformation is applied using the remaining two directions and the result is lifted back to 
11-dimensions.
If the initial D=11 background has more than 3 $U(1)$ isometries then this process can be repeated to obtain a multiparameter deformation.
In \cite{deformations} general formulas for these deformed solutions in D=11 were obtained. With the help of these, one can 
write down deformed solutions directly in D=11 without going through details of this rather lengthy calculation. To use these formulas, it does not
matter where these $U(1)$ directions lie in the geometry however the initial solution should satisfy certain conditions  
which are not too restrictive as will be seen.

Solution generating techniques were applied frequently to time-dependent backgrounds in the past (see for example \cite{old1,old2}) and actually 
in one of the first papers on the construction of S-brane solutions \cite{s1} to supergravity theories, such a method was employed \cite{s3}. 
In this paper we will study Lunin-Maldacena deformations \cite{mal2} of time-dependent solutions of D=11 supergravity, which was 
previously carried on for static M2 and M5 branes in \cite {berman}.
Using a formula derived in \cite{deformations} it is apparent that if the initial solution has zero 4-form field strength, then after 
the deformation, one gets a 3-form potential along the $U(1)$ directions which are used during the process. This is an interesting 
result since, in particular it means that if we have a cosmological solution of pure D=11 supergravity with the geometry 
$\mathbb{R}^{1+3} \times {\cal M}_7$, we can easily generalize this to a solution which has a nonzero 4-form flux along 
$\mathbb{R}^{1+3}$ using $U(1)$ directions of $\mathbb{R}^3$ which can be obtained by periodic identifications. The new  solution is electrically charged and can be interpreted as 
a generalized SM2-brane \cite{group1, pope1, pope2, s2, s3} with an arbitrary transverse space ${\cal M}_7$. 
Such geometries are quite popular in attempts to realize inflation using String/M-theory compactifications. The fact that, hyperbolic 
compactifications produce a transient accelerating phase in 4-dimensional Einstein frame was first noted in \cite{townsend}. Shortly after,
it was noticed that the transverse space could also be flat \cite{ohta} or spherical \cite{emparan} when a nonzero flux is present.
An intuitive understanding of this fact was given in \cite{emparan} using the 4-dimensional effective theory description, where the problem 
reduces to studying a potential function of scalar fields. The contribution of 4-form flux to this potential is always positive 
which enhances the amount of acceleration. Hence, it is desirable to add flux to a vacuum solution.

The plan of our paper is as follows. In the next section, we will recall the formula derived in \cite{deformations} to obtain 1-parameter
Lunin-Maldacena deformations \cite{mal2} of D=11 backgrounds. 
In section 3, we illustrate this method on some examples. Our first starting point is the 
vacuum  S-brane solution \cite{s2, townsend, ohta1} of D=11 supergravity whose 4-form field strength is zero. 
It contains several $U(1)$ directions and by using three of them, we demonstrate that one obtains the SM2-brane solution 
\cite{group1, pope1, pope2, s2, s3}
in a form which has a direct zero flux limit. We also explain the transformation of this to the familiar SM2-brane metric.
Applying the deformation further on the SM2-brane, one either gets the recently found SM2-SM2-SM5 Chern-Simons S-brane system where 
contribution of the Chern-Simons term to the field equations is non-zero \cite{chern} or the  SM2$\perp$SM2(0) intersection \cite{deger}, depending on 
which $U(1)$ directions are used during the process. Changing the initial solution to an SM5-brane  
\cite{pope1, s1} one gets either the same Chern-Simons S-brane system \cite {deger} or the SM2$\perp$SM5(1) intersection \cite{deger}. 
Successive application of the method produces solutions with more S-branes including standard intersections of 
S-branes and overlappings of Chern-Simons S-brane systems. The latter ones are new. There are also other intersections where there is 
no supersymmetric analog. In the subsection 3.2, we use this method to construct two more new solutions one of which is a new Chern-Simons 
S-brane system where there are two non-intersecting SM2-branes inside an SM5. The second one is an SM2-brane solution with a different 
transverse space than the usual one. 
In section 4, we study consequences of such deformations for accelerating 
cosmologies in (1+3)-dimensions. We show that the cosmology of a deformed solution differs from the original one 
only when we compactify on deformation directions. We also find that for the two new solutions the number of e-foldings is of order unity. 
We conclude in section 5 with some comments and possible extensions of this work.

\section{Deformations}

In this section we will explain how to generate Lunin-Maldacena deformations \cite{mal2} of an 11-dimensional background using a formula
derived in \cite{deformations}. 
The bosonic action of the 11-dimensional supergravity is
\be
S=\int d^{11}x(\sqrt{-g}R-\frac{1}{2.4!}\sqrt{-g} F^2 - \frac{1}{6}
F\wedge F \wedge A) \, ,
\ee
whose equations of motion are  
\bea\label{eins1}
R_{AB}&=&\frac{1}{2.3!}F_{ACDE}F_{B}{}^{CDE}-\frac{1}{6.4!}g_{AB}F^{2},\\
d*F&=&\frac{1}{2}F\wedge F \, .
\label{eins2}
\eea
There is also the Bianchi identity $dF=0$. This is quite a simple theory in terms of number of fields and
for a solution it is enough to specify its metric and the 4-form field strength only.

Assume that we have a solution of these equations
with the following two properties:

\noindent $(i)$ Its metric contains $I \geq 3$ commuting
isometries, which do not mix with other coordinates. 

\noindent $(ii)$ Its 4-form field strength has at most one overlapping
with these $I$ directions.

Then, one can obtain a new solution by reducing along one of these $U(1)$'s and then using
the other two for the TsT (2 T-dualities and a shift in between) transformation \cite{mal2}. Once the 3 $U(1)$ directions for the deformation 
process are decided, the choice of the reduction direction and T-duality 
directions from these
does not affect the final answer. If one uses only a single 3-torus, one gets a 1-parameter deformation. When $I >3$
one can repeat this process with different choices of 3 $U(1)$'s and obtain a multiparameter deformation. In 
\cite{deformations} general formulas for these new solutions were obtained subject to two conditions above which considerably simplify the necessary calculations. 
We will now review the formula for the 1-parameter case below.

Suppose that there are 3 $U(1)$ directions $\{x^1,x^2,x^3\}$ which possibly mix among themselves but 
not with any other coordinate in the metric and let $T$ denote the  $3 \times 3$ torus matrix that corresponds to them. The entries of this matrix 
are read from 
the metric of the original solution, i.e. $T_{mn}=g_{mn}$. Then, starting with a solution $\{F_4, g_{AB}\}$ where each term in $F_4$ has at most one 
common direction with $\{x^1,x^2,x^3\}$, after the deformation we find \cite{deformations}:
\bea \tilde{F}_4 &=& F_4 - \gamma i_{1}
i_{2} i_{3} \star_{11} F_4 + \gamma d\left(K det T \,
dx^1 \wedge dx^2 \wedge dx^3
\right), \nonumber \\
d\tilde{s}_{11}^2 &=& K^{-1/3} g_{\mu \nu}
dx^{\mu} dx^{ \nu} + K^{2/3} g_{mn} dx^m dx^n \, , 
\label{oneparameter}\\
K &=& [1 + \gamma^2 det T]^{-1} \, ,\nonumber \eea where $m,
n$ =$\{1,2, 3\}$ and $\mu,\nu$ denote the remaining directions. The new solution is given by $\{\tilde{F}_4, \tilde{g}_{AB}\}$. 
The Hodge dual $\star_{11}$ is taken in the 11-dimensions, with respect to
the undeformed metric and $i_m$ is the 
contraction with respect to the isometry direction
$\partial/\partial x^m$, i.e. $i_m \equiv i_{\partial/\partial x^m}$. Here $\gamma$ is a real deformation parameter and 
when $\gamma=0$ we go back to the original solution. From the last term in $\tilde{F}_4$ note that such a deformation 
always generates a 3-form potential along the 
deformation directions even when the original $F_4=0$, provided that $det T$ is not constant. For time-dependent solutions this term corresponds 
to the flux of a generalized SM2-brane lying along the $\{x^1,x^2,x^3\}$ directions. Its charge is proportional to the deformation parameter $\gamma$.

\section{Examples}

Now, we will study Lunin-Maldacena \cite{mal2} deformations of some cosmological solutions of D=11 supergravity using the above formula (\ref{oneparameter}).
We will first begin with the vacuum S-brane solution to clarify our method and to 
establish its connection to SM2-brane and Chern-Simons S-brane system \cite{chern} through this deformation.
Repeated usage of this method gives rise to a large number of configurations with several S-branes that can be divided into 3 classes.
The first set contains standard intersections 
of S-branes \cite{deger} which have supersymmetric analogs. In the second one we have configurations where each S-brane pair makes a 
standard intersection but overall intersection has no supersymmetric analog, which are different than those studied in \cite{nonstandard} since 
brane charges are  independent. The last category consists of overlappings between Chern-Simons S-brane systems and 
S-branes. Solutions in the first two groups can be constructed using intersection rules found in \cite{deger}, whereas those in the third one 
are new. In part 3.2, we will use this deformation to construct two additional new solutions. The first one is a new Chern-Simons S-brane 
system and the second one is an SM2 brane with a different transverse space.

\subsection{S-branes}

The vacuum S-brane solution \cite{s2, townsend, ohta1} of D=11 supergravity is given as:
\bea
\nonumber
ds^2 &=&  e^{2\l ( t-t_1)/3} (dx_1^2+dx_2^2+dx_3^2) + e^{-\l ( t -t_1)/3}\sum_{i=1}^k e^{2(b_it-c_i)} d\theta_i^2 \\
&+&  e^{-\l (t -t_1)/3} e^{2(b_0 t-c_0)} G_{n,\sigma}^{-\frac{n}{n-1}}(-dt^2 +  G_{n,\sigma} d \Sigma_{n,\sigma}^2) \, , \nonumber \\
F_4 &=& 0 \, ,
\label{vacuum}
\eea
where $d\Sigma_{n,\sigma}^2$ is the metric on the $n$-dimensional unit sphere, unit hyperbola or flat space and
\be
G_{n,\sigma} =
\left\{
\begin{array}{ccc}
m^{-2}\sinh^2\left[(n-1)m\,(t-t_0)\right],
 \,\,\, &\sigma=-1& \,\,\, \textrm{(hyperbola)}, \\
m^{-2}\cosh^2\left[(n-1)m\,(t-t_0)\right],
 \,\,\, &\sigma=1& \,\,\, \textrm{(sphere)}, \\
\exp[{2(n-1)m\,(t-t_0)}], \,\,\,  &\sigma=0& 
\,\,\, \textrm{(flat)},
\end{array}
\right. 
\label{G}
\ee
with $k+n=7$ and $ n\geq 2$. Constants satisfy
\bea
\nonumber
b_0 t - c_0 &=& -\frac{1}{n-1}\sum_{i=1}^k b_i t + \frac{1}{n-1}\sum_{i=1}^k c_i \, , \\
2 n(n-1) m^2 &=& \frac{2}{n-1} \left(\sum_{i=1}^k b_i\right)^2 + 2\sum_{i=1}^k b_i^2 + \l^2 \, . 
\label{restriction}
\eea
Here, we took the exponentials multiplying $\{x_1,x_2,x_3\}$ directions the same since after the deformation we want to have 
a homogeneous 3-dimensional space which will correspond to the worldvolume of an SM2.  We can set one of the constants 
$\{\l, b_1,..., b_k\}$ to 1 by a rescaling and one of the integration constants 
$\{t_0, t_1, c_1, ..., c_k\}$ to zero by a shift in the time coordinate. 
Since there is no mixing in the metric and the 
4-form is zero, we can use any 3 
from $x$ or $\theta$ coordinates for the deformation by assuming that they are periodic. Choosing deformation 
directions as $\{x_1, x_2, x_3\}$, we see that
the $3 \times 3$ torus matrix $T$ is diagonal with $det T= e^{2\l ( t - t_1)}$ and applying  (\ref{oneparameter}) to 
the vacuum S-brane solution (\ref{vacuum}) we find:
\bea
d\tilde{s}^2 &=&  e^{2\l ( t-t_1)/3} (1+ \g^2 e^{2\l ( t - t_1)})^{-2/3}
(dx_1^2+dx_2^2+dx_3^2) \nonumber \\
&+&  (1+ \g^2 e^{2\l ( t - t_1)})^{1/3}e^{-\l(t-t_1)/3} \left[\sum_{i=1}^k e^{2(b_it-c_i)} d\theta_i^2
+ e^{2(b_0 t-c_0)} G_{n,\sigma}^{-\frac{n}{n-1}}(-dt^2 +  G_{n,\sigma} d \Sigma_{n,\sigma}^2)\right] \nonumber \\
\tilde{F}_4 &=& \g d[e^{2\l (t - t_1)} (1+ \g^2 e^{2\l(t - t_1)})^{-1}  dx_1 \wedge dx_2 \wedge dx_3] \, , 
\label{sm2}
\eea
where $\g$ is the deformation parameter and when $\g=0$ we go back to the vacuum S-brane 
solution (\ref{vacuum}). Note also that even though we started with a solution with no 4-form field, after the deformation we have a solution with 
$\tilde{F}_4 \neq 0$. This is an SM2-brane solution located at $\{x_1,x_2,x_3\}$, however its metric is not in the familiar form which 
contains $\cosh$ functions.  To understand the relation, we scale \{$x_1,x_2,x_3\}$ coordinates with $\g^{-1/3}$ and all other coordinates 
and constants $\{m, \l, b_1,...,b_k\}$ with $\g^{1/6}$ in (\ref{vacuum}) before performing the deformation, which makes 
$\g$ disappear in $K$. Then, deforming this rescaled metric using $\{x_1,x_2,x_3\}$  directions we get:
\bea
\nonumber
d\tilde{s}\, '^{\, 2} &=&  H^{-1/3}(dx_1^2+dx_2^2+dx_3^2) 
+ H^{1/6} [ \sum_{i=1}^k e^{2(b_it-c_i)} d\theta_i^2  
+ e^{2(b_0 t-c_0)} G_{n,\sigma}^{-\frac{n}{n-1}}(G_{n,\sigma} d \Sigma_{n,\sigma}^2 - dt^2 )]  \\
\tilde{F}_4 ' &=& q \l  H^{-1}  dt \wedge dx_1 \wedge dx_2 \wedge dx_3  \, , \nonumber \\
\label{sbrane}
H &\equiv& q^2\cosh^2 \l (t-t_1) \hspace{0.5cm} , \hspace{0.5cm} q \equiv 2\g \, ,
\eea
which is the standard SM2-brane solution \cite{group1, pope1, pope2, s2, s3} with $k$-smearings whose transverse space is of the form 
$\mathbb{R}_1 \times ...\times \mathbb{R}_k \times \Sigma_{n,\sigma}$. Note that $\g \rightarrow 0$ is not well-defined anymore, i.e., this solution is valid only when 
$q \neq 0$.
This  analysis clarifies the passage from the SM2-brane solution (\ref{sm2}) to the vacuum S-brane solution (\ref{vacuum}).

From this point it is possible to continue with more deformations since there are $(k+3)$ appropriate coordinates in the initial solution
(\ref{vacuum}). We have two options which are consistent with our rules: either 
we choose one $U(1)$ direction
from the worldvolume of SM2 and two from outside or we choose all of them transverse to the SM2.
For the first choice, if we take $\{x_3,\theta_1, \theta_2\}$ directions for deforming solution given in (\ref{sbrane}), 
then this adds another SM2 along 
these directions and we get the standard SM2$\perp$SM2(0) intersection \cite{deger} where the worldvolume of the second SM2 is inhomogeneous 
with some exponentials of time which is due to the choice that we made for homogeneous directions in our initial vacuum  (\ref{vacuum}). 
For the latter, without loss of generality
let us use $\{\theta_1, \theta_2, \theta_3\}$ coordinates to deform the solution (\ref{sbrane}). Since these do not overlap with
any of the directions of the 4-form field strength of (\ref{sbrane}) and they do not mix with any coordinate 
in the metric, we are allowed to use the deformation formula (\ref{oneparameter}).
Again the $3 \times 3$ torus matrix $T$ is diagonal and after
the deformation of the SM2 solution (\ref{sbrane}) with the parameter $\g_1$ we find:
\bea 
\label{deform}
d\hat{s}_{11}^2 &=& K^{2/3} H^{1/6} \sum_{i=1}^3 e^{2(b_it-c_i)} d\theta_i^2 
+ K^{-1/3}H^{1/6} \sum_{i=4}^k e^{2(b_it-c_i)} d\theta_i^2 \\
&+& K^{-1/3}H^{-1/3}(dx_1^2+dx_2^2+dx_3^2) +
K^{-1/3} H^{1/6} G_{n,\sigma}^{-\frac{n}{n-1}} e^{2(b_0t-c_0)} \left[-  dt^2 
+  G_{n,\sigma}  d\Sigma_{n,\sigma}^2 \right] \, , \nonumber \\
K &=& [1+ \gamma_1^2 q \cosh \l (t-t_1) e^{2(bt - c)}]^{-1} \,\, , \,\, b \equiv (b_1+b_2+b_3) \,\, , \,\, c= (c_1+c_2+c_3) \, ,
\nonumber \\
\nonumber
\hat{F}_4 &=& \tilde{F}'_4 - \gamma_1 q \l  \textrm{Vol}(\theta_k) \wedge \textrm{Vol}(\Sigma_{n,\sigma}) \\
&+& 
\nonumber 
\frac{\gamma_1 q \cosh \l (t-t_1) [\l \tanh \l (t-t_1) + 2b]e^{2(bt-c)}}{[1+\gamma_1^2 q \cosh \l (t-t_1)e^{2(bt - c)}]^2} dt \wedge d\theta_1
\wedge d\theta_2 \wedge d\theta_3 \, .
\eea
This is nothing but a slight generalization of the solution given in \cite{chern}
which was previously obtained by directly solving the field equations (\ref{eins1}) and (\ref{eins2}). It is more general because 
exponentials in front of the $\{\theta_1, \theta_2, \theta_3\}$ coordinates in the metric are not all equal
and it is possible to have two smearings instead of one. 
Moreover, in  \cite{chern} the constant $\gamma_1$ does not 
appear explicitly, and hence it does not reduce to the single SM2-brane (\ref{sbrane}) by setting a constant to zero unlike the solution above.
If we take 
constants as $b_1=b_2=b_3 \equiv d/6$ and
$c_1=c_2=c_3 \equiv t_2/6$  so that the second SM2 has a homogeneous worldvolume and $n=4$, then after some further redefinitions 
of constants and rescaling of coordinates\footnote {To go from the above solution 
(\ref{deform}) to the one found in \cite{chern} first set $\l=1$ and define $\gamma_1^2qe^{-t_2} \rightarrow e^{-t_2}$. Then perform the 
changes $\theta_i \rightarrow \theta_i \gamma_1^{1/3}, \, q \rightarrow -q\gamma_1^{-1}, \, x_i \rightarrow x_i\gamma_1^{-1/3}, \, i=1,2,3$.
After these, the constant $\gamma_1$ disappears.} two solutions agree completely.
Looking at $\hat{F}_4$ we see that this new solution describes two SM2-branes located at $\{x_1, x_2, x_3\}$ and 
$\{\theta_1, \theta_2, \theta_3\}$ and an SM5-brane at $ \{x_1, x_2, x_3, \theta_1, \theta_2, \theta_3\}$.
Note that $\hat{F}_4 \wedge \hat{F}_4 \neq 0$ and therefore the contribution of the Chern-Simons term to the 
field equations (\ref{eins2}) is non-zero \cite{chern}. 

From the deformation formula (\ref{oneparameter}) we see that there is no way to obtain the SM5-brane solution \cite{pope1, s1} from any 
vacuum solution. 
However, we can start directly with the SM5-brane solution. Since the 4-form field of an SM5-brane lies along the transverse space
there are two options for deformations that are compatible with our rules: Either all 3 are chosen from the worldvolume of SM5 or two are chosen from the worldvolume and 
one from the outside. In the first case, after applying (\ref{oneparameter}) one gets again (\ref{deform})  after some obvious choices of 
constants. Whereas, from the latter starting from an SM5 with inhomogeneous worldvolume and 1-smearing
one obtains the standard SM2$\perp$SM5(1) intersection \cite{deger}. 
Thus, we have all the standard double intersections between SM2 and SM5 branes \cite{deger} except SM5$\perp$SM5(3).

Now, a large number of S-brane configurations can be constructed by  applying
more deformations that are compatible with our conditions. To increase this number one can use 
SM2, SM5 and SM5$\perp$SM5(3) as a basis and systematically perform deformations. In finding the list of resulting configurations
it is enough to remember the following set of rules about positions of available deformation directions:

SM2 $\xrightarrow{\rm{2 \,\, transverse}}$ \,\,\, SM2$\perp$SM2(0) \, , \, SM2 $\xrightarrow{\rm{3 \,\, transverse}}$ \,\,\, CSS

SM5 $\xrightarrow{\rm{2 \,\, worldvolume}}$ SM2$\perp$SM5(1) \, , \, SM5 $\xrightarrow{\rm{3 \,\, worldvolume}}$ CSS

\noindent where CSS stands for the Chern-Simons S-brane system in which there are 2 non-intersecting SM2-branes inside an SM5. Of course, when 
there are 
more than one brane in the initial system these two rules should be used simultaneously. In this way, we can get all the standard S-brane 
intersections listed in \cite{deger} which have static supersymmetric analogs. There
are also intersections where each S-brane pair makes a standard intersection but overall intersection has no supersymmetric
analog, however their construction still follows intersection rules found in \cite{deger}.
Besides these, overlappings between CSS systems and CSS systems with extra S-branes are allowed which are new in the S-brane literature.
For example, consider the SM2$\perp$SM2(0) intersection that we mentioned above where SM2's are located at $\{x_1,x_2,x_3\}$ and 
$\{x_3,\theta_1,\theta_2\}$. If we deform this using $\{\th_3,\th_4,\th_5\}$ we get an overlapping of two CSS systems 
(\ref{deform}) where there is an additional SM2  
at $\{\th_3,\th_4,\th_5\}$ and two SM5's are located at $\{x_1,x_2,x_3,\th_3,\th_4,\th_5\}$ and $\{x_3,\th_1, ..., \th_5\}$. Instead of this,
if we use $\{\th_2,\th_3,\th_4\}$  then we find a CSS system with an additional SM2 at $\{\theta_2, 
\theta_3,\theta_4\}$ where the SM5-brane is located at $\{x_1,x_2,x_3,\theta_2, 
\theta_3,\theta_4\}$. Similarly, 
using $\{x_3,\th_2,\th_3\}$ we get SM2$\perp$SM2$\perp$SM2(0) intersection \cite{deger}, whereas 
$\{x_2,\th_2,\th_3\}$ gives another SM2$\perp$SM2$\perp$SM2(-1) solution where each pair has one common direction but there is no overall common 
intersection where -1 indicates this fact. The last one is different than the solution found in \cite{nonstandard} since SM2-brane charges are independent.

\subsection{Two New Solutions}

Of course, it is straightforward to generate additional new solutions using this mechanism. For example, deforming 
the power-law solution given in \cite{freund} (see also \cite{pope2, flux}) we arrive at: 
\bea
\label{newchern}
d\tilde{s}^2 &=&  (1 + \gamma^2 \alpha_1^3 t^{-2/7})^{-2/3} 
\alpha_1 t^{-2/21}(dx_1^2 +dx_2^2 + dx_3^2) \\
\nonumber
&+& (1 + \gamma^2 \alpha_1^3 t^{-2/7})^{1/3} \left[ \alpha_1 t^{-2/21} (dx_4^2 + ... +dx_7^2) + \alpha_2 t^{-2/3} d\Sigma_ {3,-1} ^2 
- \alpha_3 t^{-8/3}dt^2 \right] \, , \\
\nonumber
\tilde{F}_4 &=& \frac{\lambda t^{-2}}{2}  dt \wedge {\rm Vol} (\Sigma_{3,-1}) + \gamma d \left[\frac{\alpha_1^3 t^{-2/7}}
{1+\gamma^2\alpha_1^3t^{-2/7}} dx_1 \wedge dx_2 \wedge dx_3 \right] 
-  \frac{\gamma \lambda \alpha_2^{-3}}{2} 
dx_4 \wedge \cdots \wedge dx_7 \, ,
\eea
where constants are fixed as $(\alpha_1)^{21}= 27\lambda^4/224$, $(\alpha_2)^3=2/(7\lambda^2)$ and $\alpha_3 =(\alpha_1)^7 (\alpha_2)^3$.
This is another SM2-SM2-SM5 Chern-Simons system for which $\tilde{F}_4 \wedge \tilde{F}_4 \neq 0$.
The initial SM2-brane has hyperbolic worldvolume $\Sigma_ {3,-1}$ and can be called a  flux SM6-brane too \cite{flux}.
The other SM2 is located at $\{x_1,x_2,x_3\}$ and SM5 worldvolume contains both of them.

As a second example we deform the vacuum solution found in \cite{product} and find:
\bea
\label{sproduct}
d\tilde{s}^2 &=& e^{2\lambda t/3}(1 + \gamma^2 e^{2 \lambda t})^{-2/3} (dx_1^2+dx_2^2+dx_3^2)  \\
&+& 
(1 + \gamma^2 e^{2 \lambda t})^{1/3} e^{-\lambda t/3} G_{7,\sigma}^{-7/6} (-e^{\beta/3} dt^2 +  G_{7,\sigma} \sum_{i=1}^n
e^{\beta_i/3} d\Sigma^2_{m_i,\sigma}) \, , \nonumber
\\
\nonumber
\tilde{F}_4 &=& \frac{ 2 \gamma \lambda e^{2 \lambda t}}{(1 + \gamma^2 e^{2\lambda t})^2} dt \wedge dx_1 \wedge dx_2 \wedge dx_3 \, ,
\eea 
where the $\Sigma_{m_i,\sigma}$'s are $m_i\geq 2$ dimensional spaces
with the same type of constant curvature specified by $\sigma$  and $\sum_{i=1}^n m_i =7$.
This represents an SM2-brane located at $\{x_1,x_2,x_3\}$ with a transverse space of the form 
$M_{m_1,\sigma} \times \cdots \times M_{m_n,\sigma}$.
The function $G_{7,\sigma}$ is given by (\ref{G}) with $m  = \lambda /\sqrt{84}$ 
and warping constants $\beta_i$ and $\beta$ are determined as 
\be
\beta_i = \frac12 \ln \left[ \frac{6}{ (m_i-1)}
\prod_{j=1}^n \left( \frac{m_i-1}{m_j-1} \right)^{m_j} \right], \,\,\,\,
\beta = \sum_{i=1}^n m_i \beta_i= \frac{7}{2} \ln \left[6
\prod_{i=1}^n (m_i-1)^{-m_i/7} \right] \, .
\ee
When the transverse space is only one piece (that is, $n=1$), all $\beta_i$'s and $\beta$ become zero and the above solution
reduces to the usual SM2-brane (\ref{sm2}) with no smearings. If desired, this again can be put into a form where $\cosh$ functions appear 
in the metric as we did above in (\ref{sbrane}), which replaces the transverse part of (\ref{sbrane}) with the above product space structure.

\section{Accelerating Cosmologies}

Hyperbolic compactifications to 4-dimensions may lead to a short period of accelerating cosmologies \cite{townsend}. 
Unfortunately, in all the examples studied so far, such as \cite{townsend, ohta, emparan, product, chern},
the amount of e-foldings is only of order 1 and hence these are not useful for explaining early time inflation.
Yet, such solutions might be relevant for the presently observed acceleration of our universe \cite{present}.

After compactification from D=11 to (1+3)-dimensions, the 4-dimensional part of all the above S-brane solutions in the Einstein 
frame has the form:
\be
ds_E^2=-S^6\,dt^2+S^2\,ds_{M_3}^2 ,
\label{einsteinframe}
\ee
where $S$ is some function of time that depends on the solution and $M_3$ is a three dimensional homogeneous space. This universal structure 
is due to a particular property of these solutions.
Namely, the function in front of the time coordinate in the metric is given as multiplication of powers of other functions appearing
in the metric where powers are the dimensions of spaces that these functions multiply. Now, the proper time is given by $d\t=S^3dt$ and 
the expansion and acceleration parameters can be found respectively as
\be
H=S^{-1}\fr{dS}{d\t}=S^{-4}\fr{dS}{dt},\hspace{1cm}  a= \fr{d^2S}{d\t^2}=-\fr{1}{2}S^{-3}\fr{d^2}{dt^2}S^{-2}.
\label{acceleration}
\ee
An accelerating phase requires $H>0$ and $a>0$. 

The effect of the deformation (\ref{oneparameter}) on a metric is to bring factors of $K$. If after the  deformation we 
compactify on a (1+3)-dimensional space whose spatial part $M_3$  was not used for the deformation, then it is immediately seen that in the 
Einstein frame (\ref{einsteinframe}) the factor $K$  does not appear in the function $S$. Hence, the cosmology
of compactication on $(t, M_3)$ will be the same before and after the deformation. On the other hand, if we use three coordinates of $M_3$ 
for the deformation, then we get a factor of $K^{-1/4}$ in the $S$ function and the cosmology will now be different.
This is actually not a surprise, since the main effect of such a deformation is to
produce a 3-form potential along the deformation directions (possibly with some additional fluxes) which will change 
the 4-dimensional cosmology only if we compactify
on these coordinates. This argument, 
together with the fact that we want $M_3$ to be a homogeneous space imply that
increasing the number of standard intersections \cite{deger} will not improve the amount of e-foldings as was explicitly checked for double intersections 
in \cite{product}. 

The above discussion shows that for the SM2-brane solution (\ref{sm2}) only compactification on $\{t, x_1,x_2,x_3\}$ may lead to a result 
different than the vacuum (\ref{vacuum}). In this case the $S$ function is given as
\be
S= e^{-\lambda t/4}(1 + \g ^2 e^{2\lambda t})^{1/4} e^{-\frac{(b_1+...+b_k)t}{2(n-1)}} e^{\frac{(c_1+...+c_k)}{2(n-1)}}
G^{-\frac{n}{4(n-1)}}_{n,\sigma} \, , 
\label{scale}
\ee
where $k+n=7$ and $k \leq 5$. The enhancement of the 4-form flux on acceleration \cite{emparan} can be clearly seen in this example. 
For the vacuum case ($\gamma=0$) an accelerating phase happens only when $\sigma=-1$ \cite{townsend}  whereas when $\gamma \neq 0$   
the expansion factor gets slightly bigger for $\sigma=-1$ and acceleration
exists  also for $\sigma=0$ and $\sigma=1$ \cite{emparan, ohta, product}.
For $\gamma \neq 0$ previously only $k=0$ \cite{ohta} and $k=1$ \cite{product} cases were analyzed explicitly. We found that increasing the number of 
flat product spaces does not lead to a significant modification and still the acceleration is of order 1.

Similarly, for the Chern-Simons S-brane system (\ref{deform}) 
compactification on $\{t, x_1, x_2, x_3\}$  will give the same answer like SM2-brane (\ref{sbrane})
as was explicitly observed in \cite{chern}. However, compactification on $\{t, \theta_1, \theta_2, \theta_3\}$ in (\ref{sbrane})
and (\ref{deform}) will give different $S$ functions and hence different cosmologies. 
For this case, choosing $b_1=b_2=b_3 \equiv d/6$ and
$c_1=c_2=c_3 \equiv t_2/6$  and we find 
\be
S = [1 + \g_1^2 q \cosh \l (t-t_1) e^{dt-t_2}]^{1/4} e^{-\frac{(n+2)(dt-t_2)}{12(n-1)}} e^{-\frac{(b_4+b_5)t -
(c_4+c_5)}{2(n-1)}} G^{-\frac{n}{4(n-1)}}_{n,\sigma} \, ,
\ee
where $2 \leq n \leq 4$. In \cite{chern} this compactification with no smearings ($n=4$)
were studied and an accelerating interval was found for each $\sigma$ . However, the amount of e-folding was again of order unity. 
When $n=3$ and $n=2$  there is a short period of acceleration too, however there is no major change in the expansion factor.

For the new Chern-Simons S-brane system that we constructed above (\ref{newchern}),
after the compactification we get a different cosmology from the original one only if we compactify on $\{t, x_1,x_2,x_3\}$. 
In this case we have
\be
S= (1 + \gamma^2 t^{-2/7})^{1/4} \, t^{-9/14} \, .
\ee
However, from (\ref{acceleration}) we find that acceleration is always negative with or without deformation.
Even though, we have flux along $\{t, x_1,x_2,x_3\}$ and some part of the transverse space is hyperbolic
we do not get an accelerating phase.

Finally, for the new SM2 solution with product transverse space (\ref{sproduct}), if we compactify on $\{t, x_1,x_2,x_3\}$
the $S$ function is
given as
\be
S =  e^{\beta/12} e^{-\lambda t/4} (1+\gamma^2 \alpha_1^3 e^{2\lambda t})^{1/4} 
G_{7,\sigma} ^{-7/24} \, ,
\ee
whose form is almost identical with the SM2-brane case (\ref{scale}). Hence, it immediately follows that when $\gamma = 0$ 
there is acceleration only for $\sigma =-1$ \cite{product}. However,  it also occurs 
for $\sigma=1$ and $\sigma=0$ after the 4-form becomes nonvanishing, alas only of order 1.

\section{Conclusions}

In this paper we looked at applications of  Lunin-Maldacena deformations \cite{mal2} to cosmological solutions of D=11 supergravity. 
The method becomes especially useful if we have a solution of pure Einstein equations which has an $\mathbb{R}^3$ part in its geometry. 
Then, using the deformation this can be generalized to a solution with a 3-form potential along these directions.
To realize inflation is a big challenge for String/M-theory and compactifications with different transverse space geometries and fluxes
is a promising way to attack this puzzle. Furthermore, this also shows that to construct flat SM2-brane solutions  
with general transverse spaces, it is enough to concentrate 
only on Einstein equations (\ref{eins1}) with $F_4=0$. We hope that with this simplification it will be easier
to construct such solutions which may have better cosmological features.

As we saw, using the deformation repeatedly it is possible to obtain configurations with several S-branes some of which are new solutions. 
If we extend our basis of initial solutions to include SM2$\perp$SM2(-1) and SM2$\perp$SM5(0) intersections found in \cite{nonstandard} 
which have no supersymmetric
analogs, then we will obtain intersections between these, standard S-branes and Chern-Simons S-brane systems too.
Cosmological aspects of such solutions need further examination. It may also be interesting to apply this method to intersections of 
S-branes with p-branes \cite{mas, sp}.

We carried on our investigation using the formula (\ref{oneparameter}) derived in \cite{deformations} which makes the calculations
very simple. In fact, once $U(1)$ directions are decided it only remains to calculate a determinant.
In this formula it is assumed that 3 $U(1)$ directions have no mixing with any other coordinate 
in the metric. However, this condition can be relaxed. In \cite{deformations} formulas where mixing of these with 1 or 2 more directions 
which are not necessarily $U(1)$ are also provided. In employing (\ref{oneparameter}) we can in principle use any 3 $U(1)$'s in the geometry 
which are consistent with our rules. 
For example, if there is an $n$-dimensional maximally symmetric piece in the geometry, than its $SO(n)$ Cartan generators can be deformed, which 
will give an SM2-brane with an unconventional worldvolume. Another consequence that follows from (\ref{oneparameter}) is that 
it is not possible to obtain the SM5-brane solution from a 
vacuum or to add a single SM5 to an existing solution. Moreover, there seems no way to construct nonstandard  
SM2$\perp$SM2(-1) and SM2$\perp$SM5(0) intersections \cite{nonstandard}. For such cases, it may be necessary to use a more general 
U-duality transformation which is worth exploring.  

In the last decade an elegant way to construct cosmological solutions of D=11 supergravity, which includes SM2-branes \cite{group2} and 
their intersections \cite{group3} with flat transverse spaces, using Kac-Moody algebras have been developed 
(for a review and more refences see \cite{rev1, rev2}). It will be interesting to work out generalization of this 
approach to cover also SM5-branes and to understand the action of Lunin-Maldacena deformations in this framework.

We can of course apply our method to static solutions as well. However, deformation of a static vacuum  
does not give an M2-brane but a static SM2 \cite{flux} whose worldvolume is Euclidean. The connection between 
one of the static versions of our Chern-Simons S-brane system \cite{chern} and composite M-brane solution \cite{dyonic} was already 
noted in \cite{chern}. The composite M-brane solution \cite{dyonic} and its intersections \cite{costa1}
were obtained using U-duality and their anisotropic black generalization were also studied \cite{costa2}. We expect our approach to be  useful 
in construction of Poincar$\rm{{\acute e}}$ symmetric versions of these black brane solutions. We aim to examine these problems in the near 
future.

\

\noindent {\Large \bf Acknowledgements}

\noindent NSD is grateful to the Abdus Salam ICTP and especially to its associates scheme, where some part of
this paper was written.

\end{document}